\documentclass[11pt]{JHEP3}
\usepackage[dvips]{graphicx}
\usepackage[centertags]{amsmath}
\usepackage{amssymb}
\usepackage{amsmath}
\usepackage{afterpage}
\usepackage{slashed}
\usepackage{epsfig}
\usepackage{multicol}

\newcommand{\bear}{\begin{array}}  \newcommand{\eear}{\end{array}}
\newcommand{\bea}{\begin{eqnarray}}  \newcommand{\eea}{\end{eqnarray}}
\newcommand{\beq}{\begin{equation}}  \newcommand{\eeq}{\end{equation}}
\newcommand{\bef}{\begin{figure}}  \newcommand{\eef}{\end{figure}}
\newcommand{\bec}{\begin{center}}  \newcommand{\eec}{\end{center}}
\newcommand{\bed}{\begin{description}}  \newcommand{\eed}{\end{description}}

\title{Baryogenesis from a  right-handed 
neutrino condensate}

\author{Gabriela Barenboim\\
	Departament de F\'{\i}sica Te\`orica and IFIC, Universitat de 
Val\`encia-CSIC, E-46100, Burjassot, Spain.\\
	E-mail: \email{gabriela.barenboim@uv.es}}

\author{Javier Rasero\\
	Departament de F\'{\i}sica Te\`orica and IFIC, Universitat de 
Val\`encia-CSIC, E-46100, Burjassot, Spain.\\
	E-mail: \email{javier.rasero@uv.es}}

\preprint{}	

\abstract{We show that the baryon asymmetry of the Universe can be
generated  by  a strongly coupled 
right handed neutrino condensate which also drives  inflation.
The resulting model has only a small number of parameters, which completely
determine not only  the baryon 
asymmetry of the Universe and the mass of the right handed neutrino  
but also the inflationary phase.   This feature allows us to make
predictions that will  be tested by 
current and planned experiments. 
As compared to the usual approach 
our dynamical framework is both economical and predictive.}

\keywords{Inflation, Baryogenesis}

\begin{document}
\section{Introduction}

\label{sec:introduction}

There is nowadays an overall consensus that modern cosmology, based on the hot
Big-Bang model and general relativity, constitutes a consistent theoretical framework which agrees {\it quantitatively} with data. It describes with 
amazing precision the evolution of the Universe from the first fraction of a second onwards. Nevertheless, such an impressive framework falls short of explaining the flatness and homogeneity of space, let alone the origin of matter and structures we observe in the universe today. As a result,  no 
decent theory of the Universe lacks a judicious period of inflation, which 
wipes out the above mentioned problems.

However, despite its wide use, inflation is far from being a theory. Inflation 
is just a set of models of the very early universe which involve a  period of exponential expansion, blowing up an extremely small region to one equivalent to the current horizon size in a fraction of a second. While the detailed particle physics mechanism responsible for inflation is not known, the basic picture makes a number of predictions that have been confirmed by observation. Inflation is thus now considered part of the standard hot Big Bang cosmology.  

There are a bewildering variety of different models to realize  inflation. In most of them however, inflation is parametrized through a single scalar field that fills space  and which  is assumed to have a potential energy. 
For a scalar field the total energy density and pressure are given by
\bea
\epsilon = 1/2 \left( \dot{\phi^2} + V (\phi) \right) \\
p = 1/2 \left( \dot{\phi^2} - V (\phi) \right) 
\eea
If the field is changing slowly, so that the kinetic terms are much smaller
than the potential ones, then we have $p \simeq - \epsilon$ and thus a
component that can produce exponential expansion if it dominates the total
energy density. Successful inflation thus requieres a phase in which the potential energy dominates the energy and pressure budget for a sufficiently long time.
 
Models of inflation differ in the assumed physical significance of the field
$\phi$, which is almost universally considered as a fundamental scalar.
Although very popular, specially in particle physics where they plague 
most theories beyond the Standard Model, it is important to keep in mind that so far no fundamental scalar field has been observed. Thus, alternatives to 
a fundamental scalar have been looked for. 
Specially interesting paths were developed by  technicolour \cite{Weinberg:1979bn, Susskind:1978ms}, extended technicolour \cite{Dimopoulos:1979es, Eichten:1979ah}, walking technicolour \cite{Holdom:1981rm} and topcolour \cite{Bardeen:1989ds}. Following this path, one of us \cite{Barenboim:2008ds} has recently pointed out the possible nature of the inflationary scalar field as a 
composite of massive right handed neutrinos. 

At first, the existence of heavy right handed neutrinos, with trivial quantum numbers under the SM group, provides the simplest explanation for the origin of the neutrino mass ( massless neutrinos  go hand in hand with the absence of right-handed neutrinos). In order to make right handed neutrinos heavy, we have to allow them to develop Majorana masses, i.e. to give up the difference
between neutrino matter and anti-matter.

Right handed neutrinos do not interact  via electromagnetic, strong or weak interactions, instead they only mix with the light SM neutrinos (via the seesaw mechanism) in such a way that the observed mixture becomes massive. 
According to the simple seesaw model of mixing, the mass of the light neutrinos is of ${\cal O}\left(m_D^2/M_{RH}\right)$, where $M_{RH}$ is the mass of the heavy neutrino and $m_D$  is a typical SM Dirac mass. The mere existence of Majorana fields, induces lepton number violation processes. This feature will play a fundamental role in our analisis.             

As if the situation were not puzzling enough, it is remarkable the no observational presence of antimatter in the Universe. Several measurements coming from BBN, CMB and SNIa quantize this asymmetry by
\beq
\eta\equiv\frac{n_B-n_{\bar{B}}}{n_\gamma} \approx 10^{-10} ,
\eeq
where $\eta$ denotes the asymmetry between baryons $n_{B}$ and antibaryons $n_{\bar{B}}$, normalized to the number of photons $n_\gamma$. As usual, the market offers a wide array of mechanisms to address this quantity. Basically, they are based upon fullfilling the so-called Sakharov conditions, which are sufficient but not neccesary to generate dynamically this asymmetry. The different scenarios span from generation of the baryon asymmetry due to decaying GUT particles, baryogenesis produced by quarks reflections in front of Higgs bubbles during a first order electroweak phase transition or leptogenesis (for an excellent review see \cite{Riotto:1999yt}). Along a completely different track, Cohen and Kaplan \cite{Cohen:1987vi} demonstrated that the existence of spontaneous CPT violation in the theory by means of a derivative coupling of a scalar field to a baryon current permits the generation of the baryon asymmetry in equilibrium, without CP violation. In this work, we will go along with this path.

Our work is organized as follows:  section \ref{sec:right} reviews 
how the addition of a four-fermion self-coupling of the right handed neutrino, if strong enough, triggers spontaneous breaking of the lepton number and produces a Majorana mass for the right handed neutrino. The cosmological implications of the effective potential (generated at one loop level) for the condensate are analyzed in  section \ref{sec:InfDyn}.
We show in  section \ref{sec:baryogenesis} that, due to lepton number violation
processes, one can produce a net lepton number density when the inflaton decays into ordinary matter. During the electroweak phase transition, such a net lepton asymmetry is converted into a baryon one via sphalerons  proccesses. We discuss in section \ref{sec:con} the results obtained and conclude.

\section{A right handed solution for the scalar field}

\label{sec:right}

In this section, we summarize the basic features of the model under which the calculation of the baryon asymmetry will be performed. 

We would be  interested in providing a dynamical origin to the scalar field, with a vacuum expectation value close to the energy scale of inflation.  In order to do so,  an effective four fermion self-coupling of the 
right handed neutrino field  of strength $G$  will be introduced by hand.
This new interaction, should be strong enough to form a neutrino condensate that will
trigger spontaneous symmetry breaking of lepton number and produce a Majorana mass for the
right-handed neutrino. Below the cutoff scale $\Lambda$,
the high frequency modes of the right handed neutrinos can be integrated, obtaining an effective theory of a Higgs-like composite field, which mimics the inflaton. 

Such a four-fermion self interaction takes the form
\beq\label{4F}
G({\bar\nu}_R^c\nu_R)(\bar{\nu}_R\nu_R^c),
\eeq
where $G$ is the dimensionful coupling constant, $\nu_R$ is the right handed neutrino and
$\nu^c$ indicates charge conjugation.
This is an effective interaction describing the physics below the cutoff  $\Lambda$. There may be other higher dimension operators, but these will have
subdominant effects at energies substantially below the cutoff scale.

In the limit of a large $N_F$, where $N_F$ is the number of right handed neutrino flavours under the new interaction, there will be a solution to the gap equation for the dynamically induced right handed neutrino mass, \bea
m_R&&=-\frac{1}{2}G\langle\bar{\nu}_R\nu_R^c\rangle\nonumber\\
&&= -2GN_F \ \int \ \frac{d^4l}{(2\pi)^4}(-1){\rm Tr}\left( \frac{i}{{\slashed{l}} - m_R}\right).
\eea
 when
\beq\label{eq:cond}
G\Lambda^2 \geq \frac{8\pi^2}{N_F}.
\eeq

When this condition is satisfied, the theory predicts a scalar bound state with a mass of order $m_R$ (to 
leading order in 1/$N_F$). This is a standard result quoted for the Nambu-Jona-Lasinio model. It
is important to stress that this bound state is a physical observable boson. 

This physical particle is a bound state of $\bar{\nu}_R\nu_R^c$, arising by the attractive four-fermion
interaction at the scale $\Lambda $ of equation (\ref{4F}). 
This composite-boson  $\Phi(x)=\rho(x) e^{i\frac{\phi(x)}{v}}$ is a complex field, with $\rho(x)$ its radial part, $\phi(x)$ the phase field and $v$ an energy scale we will identify with a vacuum expectation value(vev). This parametrization shows that the right number of the degrees of freedom is kept after 
right handed neutrino condensation at scales below $\Lambda$.

In terms of the new particle, we can rewrite equation (\ref{4F}) as 
\beq\label{eq:Yuk1}
g_o\left(\bar{\nu}_R^c\nu_R\Phi+{\rm h.c.}\right)- m_0^2\Phi^\dagger\Phi.
\eeq
Notice that the new effective scalar field does not have a kinetic term, and  it reproduces  the four fermion vertex as an induced interaction when integrated out, with the identification
\beq
G=\frac{g_o^2}{m_o^2}.
\eeq

To study the low-energy dynamics, we use the renormalization group to define effective low-energy 
couplings. This way, the running couplings at the scale $\mu $ are defined by integrating out all 
momentum-space degrees of freedom with momenta greater than $\mu$. As we run down from the scale
$\Lambda $ downward in energy, all the possible couplings consistent with symmetries will be
generated. However, it is expected that at scales below $\Lambda $, the theory can be parametrized
by an effective Lagrangian which contains only ``relevant'' operators, with canonical mass
dimension of four or less.  In our case this means that the scalar field
develops induced, fully gauged-invariant, kinetic terms and quartic term self-interactions from loop corrections, giving the renormalized lagrangian :
\beq 
{\cal L} = {\cal L}_\Phi +{\cal L}_{SM}
\eeq
with 
\beq
{\cal L}_\Phi=Z \partial_\mu\Phi \partial^\mu\Phi^\dagger+ g_o\left(\bar{\nu}_R^c\nu_R\Phi+{\rm h.c.}\right)- m_\Phi^2\Phi^\dagger\Phi -\lambda_0\left(\Phi^\dagger\Phi\right)^2 \label{eq:Lag} \ ,
\eeq
where 
\bea
&&Z=\frac{N_F \; g_0^2}{(4\pi)^2}\ \ln\left(\frac{\Lambda^2}{\mu^2}\right)\\
&&m_\Phi^2=m_o^2 - \frac{2\;  N_F \; g_o^2}{(4\pi)^2}\left(\Lambda^2-\mu^2\right)\\
&&\lambda_o=\frac{ 2 \; N_F \; g_o^4}{(4\pi)^2}\ln\left(\frac{\Lambda^2}{\mu^2}\right)\label{eq:lambd}.
\eea
and ${\cal L}_{SM} $ the standard model (SM) Lagrangian which contains, among others, a Dirac-mass term for the 
neutrino. Such a term, which couples  $\nu_R $ to the left handed $SU(2)$ 
doublet neutrino, allows  to identify the heavy SM singlet belonging to a $N_F$-dimensional
supermultiplet of the new interaction with a right handed neutrino field. It also lets us recognize its phase as lepton number.

The fact that the theory is derived from an effective four-Fermi interaction is manifested in relations (\ref{eq:Lag} - \ref{eq:lambd}) since the running couplings approach the corresponding bare couplings as $\mu \rightarrow \Lambda$.

A Lagrangian with a canonical kinetic term can be obtained by rescaling the scalar field  $\Phi  \longrightarrow \Phi/\sqrt{Z_\Phi} $ 
to get
\beq
{\cal L}_\Phi= \partial_\mu\Phi \partial^\mu\Phi^\dagger+ g\left(\bar{\nu}_R^c\nu_R\Phi+{\rm h.c.}\right)- V(\Phi).
\eeq
In adittion, one can express the theory in terms of physical quantities by means of the redefinition of the bare parameters 
\bea
g=\frac{g_o}{\sqrt{Z}}\\
m^2=\frac{m_\Phi^2}{\sqrt{Z}}\\
\lambda=\frac{\lambda_o}{Z^2}.
\eea

Once this is done, the potential for the scalar field is given by
\beq\label{eq:Potential}
V(\Phi)=m^2\Phi^\dagger\Phi+\lambda\left(\Phi^\dagger\Phi\right)^2.
\eeq

This potential involves only the radial component of the scalar field, i.e. 
it is symmetric under a global U(1) phase transformation (lepton number). Therefore, if the scalar field acquires a vacuum expectation value 
\bea
v = \sqrt{-\frac{m^2}{\lambda}},
\eea
breaking spontaneously the U(1) symmetry, the phase field would become a Goldsotone boson, massless at 
every level in pertubation theory. 

However, at energies close to Planck scale, it is expected that any global U(1) symmetry will be broken due to the black-hole dynamics which induces low energy effective operators that do not conserve global charges, such as lepton/baryon number \cite{Holman:1992us}. 
Thus, we can parametrize the explicit symmetry breaking terms by
adding to our Lagrangian the lowest dimension symmetry breaking term that can be constructed out of right handed neutrino fields, \footnote{This set up introduces, just below the Planck scale, two dimension six 
operators for $\nu_R$ but assumes the absence of the usual dimension three Majorana mass term. This may
sound unnatural, however in string theory -- our only consistent description of Planck scale physics --,
non-generic effective actions below the Planck scale are the natural expectation.} i.e.
\beq\label{4Fviolating}
G'\left[(\bar{\nu}^c_R\nu_R)^2+(\bar{\nu}_R\nu_R^c)^2\right].
\eeq
This term introduces another unknown high enegy scale $\Lambda'$, which is inversely proportional  to $G'$ , $G'\propto \frac{1}{\Lambda'^2 }$, and violates lepton number by four  units. 

On general grounds a small explicit breaking is expected, such that
 $\Lambda' > \Lambda$, so that one can also parametrize the 
effects of the symmetry
breaking term by means of the auxiliar scalar field from the compositness condition (\ref{eq:cond})
\beq\label{eq:Yuk2}
g'\left(\bar{\nu}_R^c\nu_R\Phi^\dagger+\bar{\nu}_R\nu_R^c\Phi\right) .
\eeq  

With the above expressons in mind, one can derive in a straightforward manner an expression for the mass of the right handed neutrino
\beq\label{eq:neumass}
m_R^2(\theta)=(g^2 + g'^2 + 2gg'\cos(2\theta))v^2,
\eeq 
where we use the dimensionless parametrization of the angular field $\frac{\phi}{v}=\theta$.

On the other hand, due to the explicitly breaking of the lepton U(1) symmetry, the $\theta$ field develops an effective potential from 1-loop corrections which reads
\beq
V(\theta)=-\frac{1}{(16\pi)^2}m^4_R(\theta)\cdot\ln\left[\frac{m_R^2(\theta)}{v^2}\right],
\eeq
leading to a non-zero mass for the $\theta$ field, which becomes now a Pseudo Nambu-Goldstone Boson (PNGB).

At this point, it is important to notice two features of this model in relation to its phenomenological behaviour.
The first one is that the (true) minimum of the potential is located at $2\theta = \pi$ and
does not vanish at it, therefore a redefinition of the potential will be needed. 
The second one is that since we are assuming a hierarchy between the spontaneous and explicit symmetry breaking scales, being the spontaneous the smallest,
i.e $\Lambda' > \Lambda$, the corresponding Yukawa couplings between the scalar field and the neutrinos will exhibit also the same hierarchy,  $g' <<< g$. 

Taking this into account, the potential for the scalar field, which will drive the inflationary dynamics, takes the following form
\beq\label{eq:NatPot}
V(\theta)=M^4\cdot(1+\cos(\theta)),
\eeq  
where $M$ is given in terms of the Yukawa couplings by 
\beq
M^4=-\frac{g^3g'v^4}{32\pi^2}\left(1+4\ln g\right).
\eeq

\FIGURE[t]{
\centering
 \epsfig{file=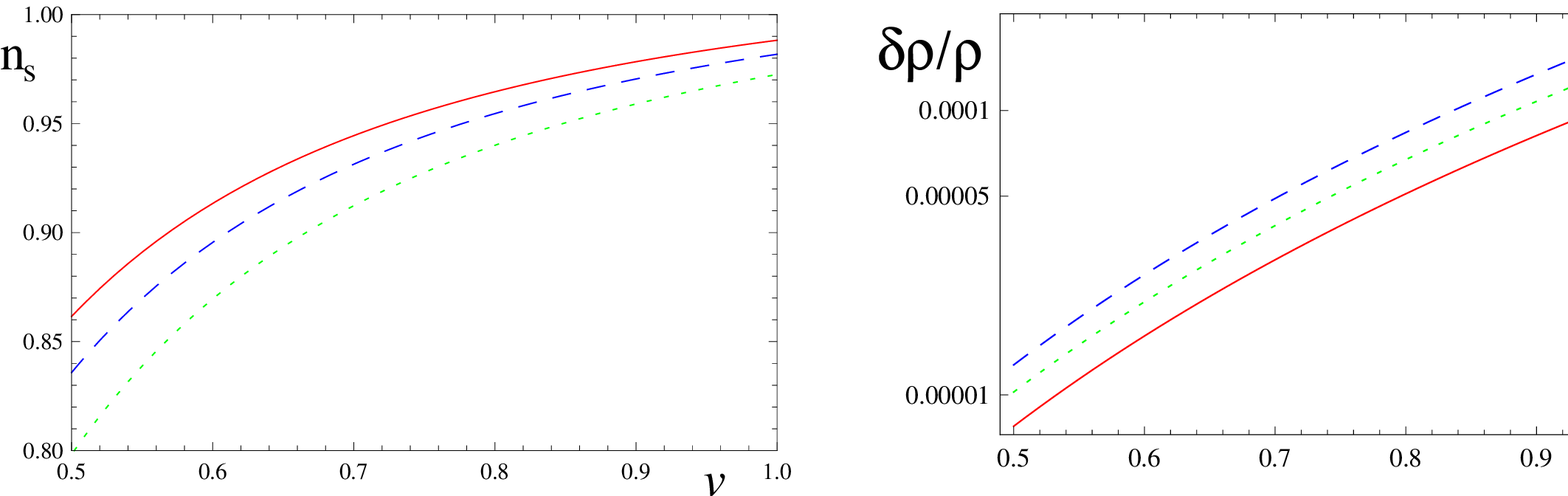, width=15cm} 
\caption{Evolution of the spectral index and the density fluctuations as a function of the spontaneously symmetry breaking scale $v$ for different couplings $(g, g')$. }
\label{fig:nsdelta}
}

The above potential is known in the literature under the name of ``Natural Inflation'' \cite{Freese:1990rb} and for certain range of its parameter space displays a potential flat enough to satisfy the inflation requirements. It also exhibits 
two widely different  energy scales: $M$ which establishes the scale of the potential  and will be related with the
energy scale at which inflation takes place
and $v$, the vacuum expectation value which will define the mass of the right handed
neutrino, that of the inflaton and the scale of spontaneous symmetry breaking, together
with the inflationary observables.

In \cite{Barenboim:2008ds}, an exhaustive analisis of the inflationary epoch has been performed to constrain the value of the Yukawa couplings and the scale of spontaneous breaking. It was found that 
\bea
0.7 {\rm M_{Pl}}\leq v \leq 0.9 {\rm M_{Pl}}\label{eq:vev}\\
\left(g^3g'\right)^{1/2} \sim 10^{-5} \label{eq:primfluc}
\eea
was needed to provide the correct scalar spectral index and size of density fluctuations.  Figure (\ref{fig:nsdelta}) shows the evolution of these observables as a function of the symmetry breaking scale for different sets of couplings $g$, $ g'$. As a result, a value of $M^4\sim (10^{16} \ {\rm GeV})^4$ must be enforced for a natural choice  of $g$. 

\section{Inflationary dynamics}
\label{sec:InfDyn}

From the potential obtained in  last section, one can reconstruct the dynamics of the inflationary field.
Tipically, almost any inflationary  transition goes trough  two recognizible periods. 
During the first one, the inflaton motion is overdamped by the huge exponential expansion of the Universe, making it evolve very slowly (slow roll phase). Owing to this proccess, the Universe  dilutes any undesearible relic and emerges extremely flat and smooth. The second epoch comprises the oscillations of the inflationay field, which gets converted into radiation, ``reheating'' the Universe. Along this phase, the inflaton mimics  nonrelativistic  matter evolution. 

During the second stage, the decay width of the inflaton can be parametrized as
\beq\label{eq:gamm}
\Gamma\simeq k_0 m_{\theta}(t),
\eeq
where $k_0$ denotes the coupling between the inflaton and relativistic matter (essentially all particles are massles, i.e. relativistic at that time)  and $m_{\theta}(t) = \sqrt{V''(\theta(t))}$ is the PNGB time variating mass, defined as the second derivative of the potential. The value of $k_0$, which sets the decay width, determines for how long the inflaton dominates the energy budget of the Universe while reheating.

The equation of motion which governs the dynamics of the inflaton can be read as
\beq
\ddot{\theta}+(3H+\Gamma)\dot{\theta}+\frac{V'(\theta)}{v^2}=0,
\eeq
where the factor $1/v^2$ arises from the parametrization for the inflaton field 
we use and $\Gamma$ is the PNGB decay width operator which takes into account the dilution of the scalar field into radiaton. 
Contrary to the traditional picture, we include this term even in the inflationary epoch.  This is the so-called ``warm inflaton'' scenario \cite{Berera:1995ie}. 
Strictly speaking one should always  include such a term. However, in  most of cases, 
$\Gamma  << H $ 
during the slow roll phase, and one can safely neglect it. 
In our case, $\Gamma$ is not so small, so we have to include it at every stage.

The evolution equations for the fields involved are well known and given by
\bea
\dot{\rho}_\phi &=& -3 H (1 + w_\phi) \rho_\phi  - \Gamma v^2\dot{\theta}^2 ,\\
\dot{\rho}_\gamma &=& -4H\rho_\gamma  + \Gamma v^2\dot{\theta}^2 \label{eq:rhorad},
\eea
where

\bea
H^2&=&\frac{8\pi}{3M_{Pl}^2}\left(\rho_\theta + \rho_\gamma\right) ,\\
\rho_\theta&=& \frac{1}{2}v^2\dot{\theta}^2 + V(\theta) ,\\
V(\theta)&=& M^4\left(1+\cos(\theta)\right) ,
\eea
with the dimensionless parametrization $\theta = \frac{\phi}{v}$.

Solving numerically this set of equations, one obtains the thermal history of the universe 
and the behaviour of the inflaton during its rolling down of the potential. The general pattern which follows from these equations is clear. The inflaton starts dominating the energy density of the Universe, only diluting away as a consequence of the expansion. Once the end of the slow-roll phase is reached, the friction term becomes dominant and converts the inflaton energy into radiation, reheating the Universe and recovering the old Big Bang picture. The point where both components cross 
depends on the the precise value of $\Gamma$. This point signals a time $ t \approx \Gamma^{-1}$ at which most of the energy stored in the inflaton field
gets converted into radiation. This feature motivates the widely used instantaneous decay approximation, which is a qualitative/easy way to analytically solve the equations of motion
and provides a picture that captures the essence of the behaviour of the inflaton during the reheating process. 

Alternatively, when one is dealing with an inflationary scheme, it is important to know the temperature reached once the inflaton has completely decayed away. This temperature, normally called the reheating temperature,
is determined by the radiation energy density generated as follows
\beq
T = \left(\frac{30\rho_\gamma}{g_*\pi^2} \right)^{1/4},
\eeq 
where $g_*$ is the number of relativistic degrees of freedom in the theory 
($\approx100$ within the  standard model). Fig (\ref{fig:temps}) shows the evolution of the temperature for different $k_0$ values.
From there it can be seen that the larger the $k_0$, the shorter the
 matter domination period at the end of the inflationary phase.

\FIGURE[t]{\epsfig{file=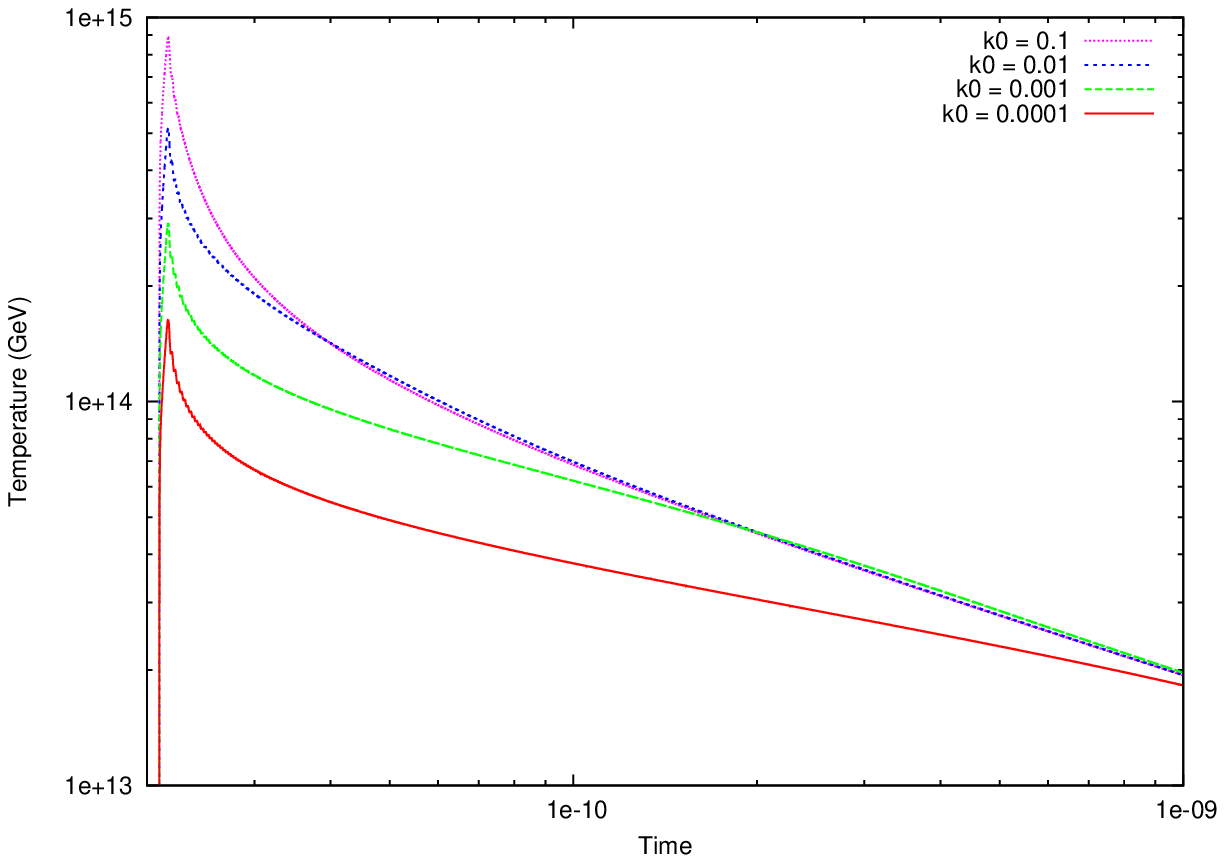} \caption{Temperature evolution depending on the reheating parameter $k_0$.} \label{fig:temps} }

Between the end of the slow roll phase and the time $t\simeq \Gamma^{-1}$ (the instantaneous decay time), 
equation (\ref{eq:rhorad})
 is dominated by the kinetic term of the $\theta$ field and thus, the temperature does not fall as in a radiation-like dominated Universe, but as $T \approx t^{-1/4}$ due to the entropy release of the 
decays. During this phase, the temperature
reaches an almost flat plateaux from the point where the energy density in the
 radiation born out of the inflaton and the energy density of the inflaton itself became
comparable up to $t\simeq \Gamma^{-1}$.
This fact can be seen clearly when $k_0= 0.001$, and $k_0 = 0.0001$ to a greater or lesser extent. 
After $t\simeq \Gamma^{-1}$, the Universe becomes radiation dominated, the expansion term dominates in eq (\ref{eq:rhorad}) and then, the temperature falls like $T \approx t^{-1/2}$.  

Tipically temperatures reached after our inflaton decayed away are of order of $10^{14} {\rm GeV}$. Such high temperatures can be problematic in models beyond the standard model, like supersymmetry, because it would lead to an overproduction of gravitinos \cite{Khlopov:1984pf}, which would have catastrophic consequences for the evolution of the Universe and specially in the formation of light elements (H, He..) at BBN \cite{Falomkin:1984eu,Khlopov:1993ye}. However,  they are perfectly acceptable in the context of the standard model.

\section{Baryon Asymmetry Calculation}

\label{sec:baryogenesis}
In this section, we calculate the baryon asymmetry generated within this model, for which
we derivatively couple the Pseudo Nambu-Goldstone boson 
to a leptonic current.
In order to obtain a non zero expectation value for the time derivative of the Goldstone field, we have included in the Lagrangian a term that soflty breaks the U(1) symmetry
explicitly as well as spontaneously. 
The above mentioned derivative coupling takes the form
\beq\label{eq:deriv}
\frac{1}{f}\partial_\mu\phi J^\mu_L,
\eeq
where $J^{\mu}_L$ is the lepton current and $f$ is associated to 
the energy scale responsible for such a term.  This sort of coupling would be only possible if, as happens in our model, lepton asymmetry is violated, otherwise the divergence of the current would vanish.
Our inflationary phase now, is just a textbook example of a second order phase transtition,
where a scalar order parameter (our phase field) evolves from one field value to 
another, as the true minimum of its effective potential changes. In the meantime, there will be a period during which the velocity of the field develops
an expectation value.

This term implies a Time Reversal and Lorentz invariance violation, which likewise will lead to a temporary violation of CPT. Even though this could scare any responsible reader, to do so locally is perfectly consistent \cite{Colladay:1996iz}.  Mild violations of CPT could have an origin in the neutrino sector \cite{Barenboim:2001ac, Kostelecky:2003cr}. 
Regarding  the possible origin of this term, when dealing with  theories near Planck scale, due to non global lepton charge conservation, the divergence of the Lepton current is non-zero, making this term suitable to appear in the Lagrangian as an effective operator.   

Regarding baryogenesis, CPT violation in the theory relaxes the Sakharov conditions for generating the baryon asymmetry dynamically. Normally, in addition to the baryon number and CP violation, one has to consider a scenario where thermal equllibrium can not be reached, since along with CPT conservation it enforces the production of a zero net baryon number. The reason is clear (recall that baryon number is an odd quantity under a CPT transformation)
\bea
<\hat{B}> &&= \mbox{Tr}\left[e^{-\beta H_{CPT}} B\right] = \mbox{Tr}\left[(CPT)e^{-\beta H_{CPT}} (CPT)^{-1} (CPT) \ \hat{B}\ (CPT)^{-1}\right] \nonumber\\
 &&= \mbox{Tr}\left[(+1)e^{-\beta H_{CPT}}\cdot(-1)\hat{B}\right] = - <\hat{B}> \ \Longrightarrow \ <\hat{B}> = 0 \, ,
\eea
where $H_{CPT}$ is CPT-conserving Hamiltonian and $\beta = 1/T$. However, the above expression no longer holds when CPT is violated in the Lagrangian and therefore, a net asymmetry can be produced even in thermal equilibrium. In addition, there is no need to break CP (or departure of equilibrium) since the  
Sahkarov conditions do not apply when CPT is violated .

We are interested in relating the inflationary scalar field with the baryon asymmetry production. Therefore, by identifying $\dot{\phi} \longrightarrow \mu$ and $\frac{v}{f} \longrightarrow\lambda$, 
a dimensionless coupling to be constrained later, the equation of motion can be re-written as 
\beq\label{eq:yoquese}
\ddot{\phi}+3H\dot{\phi}+V'(\phi) +\Gamma\dot{\phi} = -\frac{\lambda}{v}(\dot{n}_L+3Hn_L),
\eeq
where $\Gamma$ without subscript refers to the usual inflaton decay width 
into radiation, eq (\ref{eq:gamm}), and $n_L$ denotes the lepton number density. 

The above equations describe the lepton asymmetry produced in thermal
equilibrium during the inflaton slow roll down
and subsequent decay  with the interaction term  shown in eq (\ref{eq:deriv}), provided
that the rate of change of $\dot{\phi}$ is sufficiently low. If this were the case, this interaction  would shift the lepton and antilepton energy levels like a chemical potential for lepton number. (Here sufficiently low simply means that the typical time scale of lepton violating processes must be fast enough to maintain thermal equilibrium).

If thermal equilibrium cannot be reached,
one has to substitute the divergence operator by the operator that violates
lepton number. As this term gives rise to the decay of the inflaton field, one can
approximate the effect of the decay of the motion of the inflaton field due to its
lepton number violating interactions by including an extra friction term, 
proportional to the width of the lepton number violation 
\beq
\ddot{\phi}+3H\dot{\phi}+ V'(\phi)+\Gamma\dot{\phi} = - \Gamma_L\dot{\phi} \ .
\eeq
where $\Gamma_L$ is the interaction width which violates lepton charge. In this case, we differenciate both, because we are interested in the contribution of this last term to 
 the  baryon/lepton asymmetry generated. 
In general, this term should be included inside the entropy term production $\Gamma$.

Comparing both equations, we can see that the Boltzmann equation for the lepton asymmetry is given by
\beq\label{eq:Boltzm}
\dot{n}_L+3Hn_L=-\frac{1}{\lambda}\Gamma_L v \dot{\phi}.
\eeq
 
The analitic solution of this Boltzmann equation is given by
\beq
n_L(t)=n_L(t_o)-\left(\frac{a}{a_o}\right)^3\times \int_{t_o}^{t_f} \ dt \frac{1}{\lambda}\Gamma_L v \dot{\phi}
\eeq
One notices that the lepton asymmetry is just the area enclosed by the phase field 
throughout its oscillatory movement around its minimum, during which the inflaton produces leptons/antileptons for positive/negative velocities (by negative we mean velocities in the opposite direction). As this procces is modulated by the expansion of the Universe, it leads to a non zero value.

In our model, the lepton violating operator takes the form
\beq
\frac{1}{\Lambda'^2}(\bar{\nu}^c_R\nu_R)^2+h.c,
\eeq
This term comes from the explicit lepton number violation term, eq (\ref{4Fviolating}), 
with the identification $G'\equiv 1/(\Lambda')^2$, i.e, we are assuming the maximal value this
coupling constant may have. Like any  dimension six operator, ours yields a decay width of the form
\beq\label{eq:GammaL}
\Gamma_L=\lambda^2\frac{m_\theta ^5}{\Lambda'^4} \ ,
\eeq
where $m_\theta= \sqrt{V''(\theta)}$  is the inflaton mass.

\FIGURE[t]{\centering{\epsfig{file=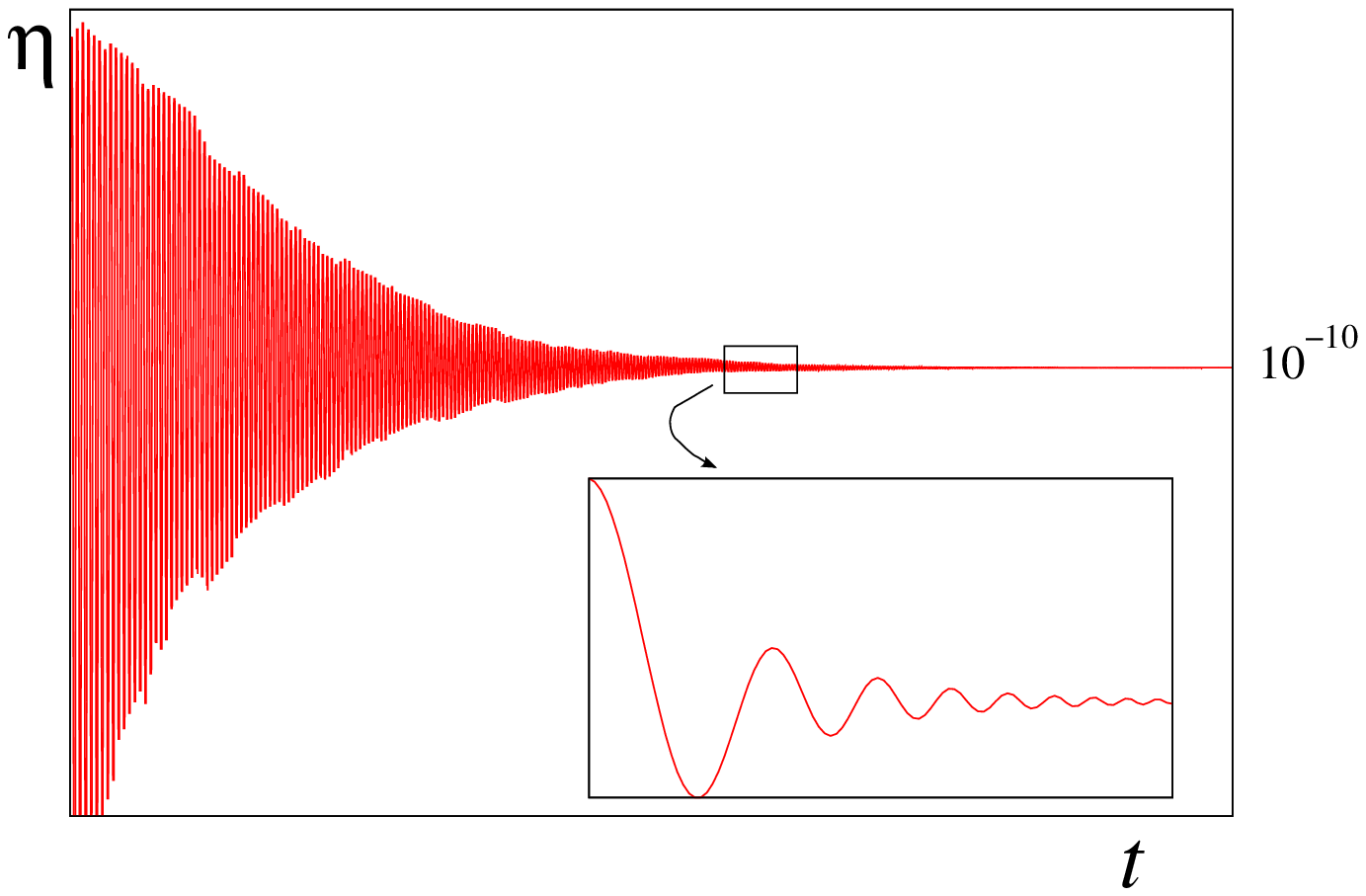,width=12cm}} \caption{Baryon asymetry evolution for $k_0 = 0.01$, $\lambda=0,01$ and $\Lambda'= 10^{16} \ {\rm GeV}$ in logarithmic scale. The amplified snapshot shows a cartoon picture of the oscillations.} \label{fig:bau} }

We are interested not in the lepton density or the baryon one, but the baryon to photon
ratio. At a temperature $T$ the photon number density is given by
\bea
n_\gamma=\frac{2\zeta(3)}{\pi^2} T^3 ,
\eea 
Thus, once we  solve numerically the equation for the lepton number asymmetry (\ref{eq:Boltzm}), we
can estimate the  baryon asymmetry $\eta = \frac{n_B}{n_\gamma}$. As we have previously mentioned, since this quantity tracks $\dot{\phi}$, one would expect a damped oscillating behaviour asymptotically reaching a final value, 
once there is no sufficient feedback to keep producing it. We show the particular feature in Fig. (\ref{fig:bau}) for the following set of values  
\bea
k_0 & = & 0.01 \ \label{eq:k} , \\
\Lambda'&\simeq&10^{16} {\rm GeV} \label{eq:L} \ ,\\
\lambda&\simeq& 0.01 \ \label{eq:l} ,
\eea
which give the experimentally observed baryon asymmetry. On the other hand, the evolution of the velocity with time will resemble the one showed, but asymptotically
 converging to zero as a consequence of the progresssive lost of kinetic energy. 

For this calculation we have taken into account that 
the lepton to baryon asymmetry convertion by the sphaleron proccess at later epochs is given by
\beq
n_B=-\frac{28}{79}n_L \ .
\eeq

Similarly, as an additional/alternative source of lepton asymmetry we could
have  used the lepton number violating operator
\beq
g'\left(\bar{\nu}_R^c\nu_R\Phi^\dagger+ {\rm h.c}\right).
\eeq

Contrary to the first one, this term is a four dimensional operator, so 
it would produce a decay width with the following form
\beq
\Gamma_L = \frac{g'^2}{8\pi}m_\theta.
\eeq

Comparing this last term with the one given in eq (\ref{eq:GammaL}), one can make an educated guess
for the value of $g'$ from the values of $\Lambda'$ and $\lambda$ needed to get the right amount 
of baryon asymmetry (eq \ref{eq:k} - \ref{eq:l}). This turns out to be
\beq
g'\simeq 10^{-7} \, .
\eeq 

The value of $g'$ can then 
be used now to determine $g$ by requiring the size of primordial
fluctuations to agree with experiment, eq (\ref{eq:primfluc}) and both together constraint the mass of the right handed neutrino $m_R$ (eq \ref{eq:neumass}). Consecuently, these parameters take the following values 
\bea
g&\simeq& 0.1, \\
m_R &\simeq &  g v  = 10^{18} \; \left(\frac{v}{M_{\mbox{Pl}}}\right) \; {\rm GeV}.
\eea
Up to an order of magnitude smaller masses for the right handed neutrino field can be obtained 
for larger values of $g'$. However, given that in the $g' < g$ regime we are forced to have 
$ g^3  g' \simeq 10^{-10}$ to provide the right size
of scalar density perturbations,  $m_R > 10^{16} ( v/ M_{\mbox{Pl}})$GeV for any choice of
fermion couplings.

With this at hand, we can already test our model. As it is well known in addition to scalar (density) perturbations, our field will also give rise to tensor (gravitational wave) perturbations. Generally, the tensor amplitude is given in terms of the tensor/scalar ratio 
\beq
r \equiv \frac{P_T}{P_R} = 16 \epsilon 
\eeq
The tensor to scalar ratio $r$ goes like  $\; g^2 g^{\prime 2} \; $
and for our model it turns out to be well below the detection sensitivity of current and (near) future experiment. 
Gravity waves are the holy grail of next generation of experiments and if found, will rule out this model.

Strictly speaking, $n_s$ is not a constant, and its dependence on the scale can be characterized by its running.  Our model predicts a very small and negative spectral index running, scaling as $g^\prime / g$. It is so negligible small  that it is essentially indistinguishable
from zero running. Small scale CMB experiments  will provide more stringent tests on the running. If these
experiments exclude a trivial (consistent with zero)  running, i.e. if they detect a strong running, our model would be ruled out.

\section{Discussion and Conclusions}

\label{sec:con}

In this paper, we have discussed the baryon asymmetry generated in an inflationary model, without a fundamental scalar field. We have showed that it is possible to obtain the observed $\eta$ value from an inflaton-like composite generated out of strongly coupled  right handed neutrinos, while at the same time agreeing with cosmological observations. 

The possibility of 
dynamically generating a scalar field, responsible not only for  breaking the symmetry
but also for giving mass to the right handed neutrino masses and 
whose decay generates the baryon asymmetry of the universe by using the CPT non-invariance of the universe during its early history makes the model
especially economic and therefore physically appealing.

The resulting model is phenomenologically tightly constrained, and can be experimentally
(dis)probed in the near future.

\subsection*{ACKNOWLEDGMENTS}
It is a pleasure to thank Joe Lykken and Oscar Vives for useful discussions. 
JR is really grateful to Simon Kiesewetter for his ineffable help with the numerical aspects. The authors 
acknowledge financial support from spanish MEC and FEDER (EC) under grant 
FPA2008-02878, and Generalitat Valenciana under the grant PROMETEO/2008/004.

\end{document}